\documentclass[10pt,letterpaper]{article}

\usepackage{graphicx}
\usepackage{opex3}
\usepackage{eepic}
\usepackage{color}

\begin{document}

\newcommand{\vect}{\vec}

\title{Nanoparticle characterization by using Tilted Laser Microscopy:\\
back scattering measurement in near field.}

\author{D. Brogioli, D. Salerno, V. Cassina, and F. Mantegazza}

\address{
Dipartimento di Medicina Sperimentale, Universit\`a degli Studi di
Milano - Bicocca, Via Cadore 48, Monza (MI) 20052, Italy. }

\date{\today}

\begin{abstract}
By using scattering in near field techniques, a microscope can be
easily turned into a device measuring static and dynamic light
scattering, very useful for the characterization of
nanoparticle dispersions. Up to now, microscopy based techniques have been
limited to forward scattering, up to a maximum of $30^{\circ}$. 
In this paper we
present a novel optical scheme that overcomes this limitation,
extending the detection range to angles larger than $90^{\circ}$
(back-scattering). Our optical scheme is
based on a microscope, a wide numerical aperture objective, and a
laser illumination, with the collimated beam positioned at a large angle
with respect to the optical axis of the objective (Tilted Laser Microscopy, TLM). 
We present here an extension of the theory of near field scattering,
which usually applies only to paraxial scattering, to our strongly out-of-axis situation. 
We tested our instrument and our calculations 
with calibrated spherical nanoparticles of several
different diameters, performing static and dynamic scattering
measurements up to $110^{\circ}$. The measured static spectra and
decay times are compatible with the Mie theory and the
diffusion coefficients provided by the Stokes-Einstein equation. The ability
of performing backscattering
measurements with this modified microscope opens the way to new applications of 
scattering in near field techniques
to the measurement of systems with strongly angle dependent scattering.
\end{abstract}


\section{Introduction.}

Optical microscopy and light scattering are widely used,
well-known techniques, often applied to the study of 
samples such as biomedical systems, colloidal suspensions, or complex fluids. 
Traditionally, the instruments for implementing optical microscopy and light scattering 
belong to two well distinct groups. 
Optical microscopes are imaging tools, that is, they provide a
deterministic mapping of an object's optical properties, collecting light
in the so-called near field\footnote{The term ``near field''
refers to the Fresnel region, as opposed to the far field Fraunhofer region,
as classically reported in classical optics textbook \cite{hecht}. It is also worth 
underlining that, in the present paper, the name ``near field'' in conjunction
with ``microscopy'' has no relation with evanescent-wave based techniques like
Scanning Near-Field Optical Microscopy (SNOM) \cite{pohl1984}, Total Internal 
Reflection Fluorescent Microscopy (TIRFM) \cite{axelrod2001}, Total Internal Reflection 
Scattering (TIRS) \cite{potenza2004}, nor Total Internal Reflection Microscopy (TIRM)
\cite{prieve1999}.}.
Conversely light scattering devices measure
statistical properties, detecting the light scattered in the far
field \cite{berne2000, van_de_hulst1981, chu2007,
pusey1985, degiorgio1980}. Accordingly, traditional light scattering methods can be
conveniently renamed as Scattering In the Far Field (SIFF) techniques.

A whole family of alternative methods have been developed for measuring the scattered light
intensity fluctuations in the near field, similarly to what is done in
microscopy. 
Several different names have been introduced to describe these specific slightly different
methods, often independently developed by different communities.
Here we will call this family: Scattering In the Near
Field (SINF) \cite{brogioli2008, brogioli2009}, as opposed to SIFF.
The community interested in laser speckles developed the Near Field Speckle (or Scattering)
(NFS) techniques, first in the homodyne version \cite{giglio2000, giglio2001} and then
in heterodyne version, Heterodyne Near Field Speckles (HNFS)
\cite{brogioli2002, brogioli_phd, ferri2004}. Similar techniques, but applied to imaging
configurations are shadowgraph \cite{wu1995,trainoff2002} and schlieren \cite{brogioli2003}.
In turn, these bear a strong similarity with Gabor's in-line holography \cite{repetto2004}.
Furthermore, the community devoted to holography have also used a similar approach for back
scattering 
detection \cite{lesaffre2006}. Finally, microscopists reported Fourier Transform Light 
Scattering \cite{ding2008} and Differential Dynamic Microscopy \cite{cerbino2008}.
However, an exaustive list of SINF techniques is out of the scope of a paper.

Some years ago, the interest in the development of a hybrid
instrument, combining optical microscopy and light scattering, led to the development of optical
systems able to record simultaneously the near and the far field
light intensities \cite{dzakpasu2004,kaplan1999,popp2001,amin2007}.
Unfortunately the extreme complexity of the experimental set-ups
limited the spread of these techniques.

SINF techniques proved to be a very simple and efficient way to
use a microscope in order to measure static light scattering \cite{giglio2000,
giglio2001, brogioli2003} or dynamic light scattering of
nanoparticles \cite{magatti2008}, depolarized dynamic light scattering
of rod-like nanoparticles \cite{brogioli2008} and biological samples
\cite{ding2008}. 
Generally the dynamics is extracted by
means of a so-called ``differential'' algorithm presented in References 
\cite{croccolo_infm2003, croccolo_amsterdam2004, croccolo2006, croccolo2006bis},
which has also been used with white light \cite{cerbino2008} and X-rays illumination
\cite{cerbino2008bis}; moreover it has been used for exctracting static data \cite{ferri2004}.
Since the algorithm involves fast grabbing of the images, a further
simplifying step is constituted by the
``exposure time dependent spectrum'' (ETDS) procedure
\cite{brogioli2008, oh2004}, which drops the requirement of a fast
acquisition.

All the above SINF experiments have been limited to small scattering
angles, up to about $30^{\circ}$, the main limitation being the
numerical aperture of the collecting objective.
By using larger numerical aperture, the maximum
detectable scattering angle can be increased, but, 
the $90^{\circ}$ limit remains inaccessible.

In this paper, we describe a novel detection scheme able to
overcome this limitation and allowing detecting the scattered light up to the
back scattering regime, (more than $90^{\circ}$).
Experimental data will be presented for scattering angles up to
$110^{\circ}$, almost four times larger than what ever reported in a previous SINF measurement. 
Potentially the method can be extended to even larger
scattering angles, approaching $180^{\circ}$.

The basic idea of the method is to illuminate the
sample with a tilted beam of coherent light, whence the name of
Tilted Laser Microscopy (TLM). Indeed, whereas in a conventional SINF
experiment a laser beam is sent through the sample along the
axis of the collecting optics, in the described apparatus
the laser beam enters into the system at a large
angle with respect to the optical axis. In the proposed optical scheme, 
the aperture of the objective and the tilt angle are chosen so that 
the transmitted and the scattered beams are both collected by
the objective. This approach is different from
the classical ``dark field'' microscopy: in dark field
the transmitted beam is deliberately removed, while in TLM
the unscattered and scattered beams overlap.

The importance of collecting the scattering at large angles
lies in the fact that the scattered light can contain details 
which are very important for nanoparticle characterization. 
For example, very small particles scatter light almost isotropically, 
while larger ones (sometimes contaminants or impurities) 
scatter more strongly in the forward direction, and so often disturb the 
detection of the nanoparticle signal at low scattering angles. 
Thus the possibility of distinguishing nanoparticles 
is directly related to the ability 
of getting information about the light scattered in the
backward direction. The proposed technique is an answer to this
problem of the low signal in nanoparticle scattering measurements. 
On the other hand, the scattering
signal frequently shows interesting features at large angles not
accessible with traditional SINF methods. Several 
scientific milestones of soft matter (fractal 
dimension of colloidal aggregates, gel/glass 
properties, liquid crystals structures, etc.) has 
been demonstrated by studying the structure 
factor of the system, up to high wave vectors.

%
\section{Scattering In the Near Field tutorial.}

\begin{figure}[h]
\begin{center}
\includegraphics{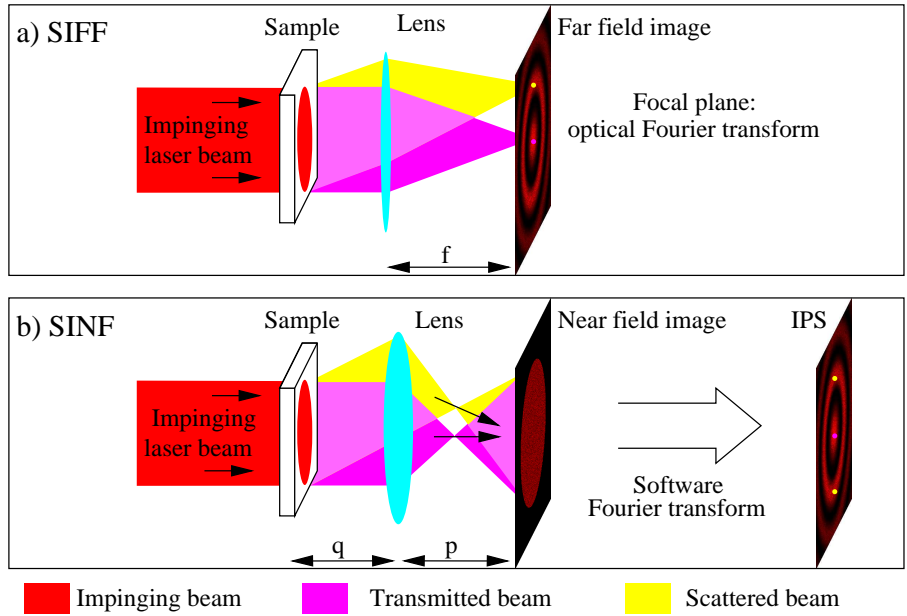}
\end{center}
\caption{Schematic description of scattering detection techniques.
The red, yellow, and magenta beams represent the impinging, scattered and transmitted beams 
respectively.
Panel a: traditional Scattering In Far Field (SIFF) technique. The scheme refers to 
a typical small angle light scattering; other SIFF techniques are based on the same principle.
The beam scattered at a given angle is focused by the lens into one point (yellow dot) on the
observation plane, where the far field image is collected. Panel b:
Scattering In Near Field (SINF) technique. The reported scheme refers to 
shadowgraph technique, and is quite similar to HNFS configuration; other
SINF techniques are based on the same principle, with slightly different
schemes. A near field image of a plane close
to the sample is formed on the observation plane. The scattered and the transmitted
beams are both interfering on the observation plane. The near field
Image Power Spectrum (IPS) is then evaluated through a Fourier transform.
The yellow beam generates two points (yellow dots) on the IPS}
\label{fig_tutorial}
\end{figure}
In this section the general ideas of Light Scattering are
briefly and quantitatively reviewed, comparing the SINF outputs with the
results of the classical SIFF technique. First, an informal and intuitive
discussion about the physics of Light Scattering is provided,
including a description of the traditional far field
techniques. Second, we discuss some 
peculiarities of the near field method, focusing on the heterodyne scheme. 
A more detailed and systematic
description of SIFF can be found in classical references
\cite{berne2000, van_de_hulst1981, chu2007, pusey1985,
degiorgio1980}, whereas rigorous presentations of SINF can be
found in \cite{brogioli2002, giglio2000, giglio2001, trainoff2002,
brogioli2003, ferri2004, croccolo2008}.

Generally speaking, a sample scatters light if its fluctuations of
concentration and temperature induce refractive index
fluctuations. In a standard light scattering experiment the
intensity of the scattered light $I\left(\vartheta, \varphi\right)$
is measured at a given scattering angle $\vartheta$ and azimuthal
angle $\varphi$. Then, for example, the scattering intensity from
spherical colloids $I\left(\vartheta, \varphi\right)$ can be
compared with the outcome of the exact Mie algorithm for determining
the particle size.

A traditional SIFF experiment basically consists in measuring the
scattered intensity at a given scattering angle selected by
placing the detector along a given direction very far from the
sample. The far field requirement is alternatively satisfied by
placing a lens just beyond the sample and collecting the light at
the lens focal plane (see Fig.~\ref{fig_tutorial}-a). The intensity
distribution thus obtained at the sensor plane exactly represents
the far field image: the light scattered at a given angle falls on a
given point $\left(\vartheta, \varphi\right)$ of the image. The
effect of the lens is essentially that of performing the optical
Fourier transform of the signal. The image (intensity
mapping) is speckled, but with a well defined average intensity
distribution, representing the time averaged scattering intensity. 
The speckled appearance is due
to the stochastic interference between the beams scattered in the
same direction by the different scatterers. Furthermore the 2D
statistical intensity distribution of the speckles is an exponentially
decaying function \cite{goodman2005}. If the sample is not static in time, the
intensity of the speckle field is fluctuating (boiling) on a time
scale depending on the scattering angle $\vartheta$. For example,
for a colloidal solution, the boiling is slow close to the forward
direction $\vartheta=0$, while it seems faster at larger angles.

In a SINF measurement, a near field image is collected either directly by a
pixilated sensor placed in front of the sample, or via an optical
imaging system (see Fig.~\ref{fig_tutorial}-b). The scheme reported 
in Fig.~\ref{fig_tutorial}-b refers to shadowgraph technique, and is
quite similar to HNFS configuration; other SINF techniques are based on
the same principle, with slightly different schemes.
The near field image consists in the 
patterns generated by interference between the overlapping scattered beam and
much more intense transmitted beam (heterodyne detection). Each
scattered beam generates a Fourier mode on the near field image;
thus the image decomposition into Fourier modes allows the determination
of the scattered beam intensities. Indeed the near field Image Power
Spectrum (IPS) is then calculated \emph{in silico} by means of a 2D
Fast Fourier Transform software. The obtained IPS is very similar to
the far field image obtained in the SIFF experiment: essentially the
same information is obtained by means of a software elaboration on the
SINF image, instead of the optical Fourier transform given by the SIFF
image.

However, some important differences between the IPS and the far
field image can be noticed.
\begin{enumerate}
\item
A single scattered beam gives rise to two points on the IPS. For
example the yellow beam of \ref{fig_tutorial}-b generates the two
points marked with the yellow dots on the IPS. In other words, the single
yellow dot shown in the far field image of Fig.~\ref{fig_tutorial}-a
corresponds to two yellow dots in the IPS of Fig.
\ref{fig_tutorial}-b. This effect is related to a fundamental property of
the Fourier transform, which is not able to distinguish between
the two symmetric beams generating a given mode within the image.
\item
The 2D statistical intensity distribution of the speckles in a
SINF image has, with very good approximation, a Gaussian distribution \cite{goodman2005}, resulting
from heterodyne superposition of the scattered and
transmitted beams. As already stated, the SIFF speckles
show a 2D exponentially decaying statistics \cite{goodman2005}.
\item
Also the dynamics is different: SINF and SIFF 
images are intrinsically related to
the CCD exposure time in different ways. While different acquisition
times in SINF correspond to a blurring of the fluctuations
\cite{brogioli2008}, in SIFF they correspond to different averages
of speckle intensities. More explicitly, while in SINF a long
exposure time reduces the power spectrum,
in the SIFF case a long exposure time produces
a less speckled appearance of the far field image.
\item
The most simple SINF schemes are heterodyne, so that they measure one of the complex
components of the radiation field, while homodyne SIFF measures only the intensity.
As consequence, a near field image carries information about the phases.
As a result, it is easy to perform velocimetry with the SINF heterodyne signal
\cite{alaimo2006}. On the other hand, this sensitivity to phases can be seen as a drawback, 
since a diffusion measurement is vulnerable to collective motions.
\item
While the SIFF signal
provides a direct measurements of the scattered intensity, the SINF signal (the Image Power spectrum)
is related to the scattering intensity via the transfer function of the system. The latter one, depending
on geometry and scattering angles, may introduce not negligible corrections.
\end{enumerate}

\section{Tilted laser methods.}

\begin{figure}[h]
\begin{center}
\includegraphics{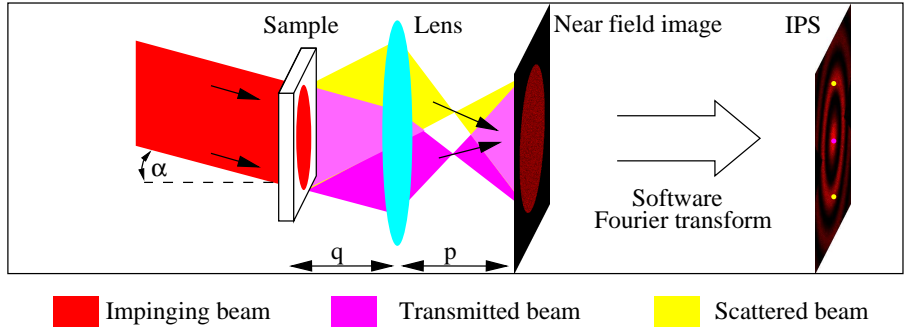}
\end{center}
\caption{Schematic description of Tilted Laser Microscopy technique.
The red, yellow, and magenta beams represent the impinging, scattered and transmitted beams 
respectively.
The impinging beam falls on the sample at angle $\alpha$ with respect to
the optical axis of the system. A near field image of a plane close to
the sample is formed on the observation plane. The near field
Image Power Spectrum (IPS) is then evaluated through a Fourier transform.
The yellow beam generates two points (yellow dots) on the IPS.}
\label{fig_tutorial_c}
\end{figure}

Figure \ref{fig_tutorial_c} schematically describes how the here proposed TLM setup works. Basically, 
the scheme is analogous to the SINF design shown in Fig.~\ref{fig_tutorial}-b,
with the difference that in the tilted scheme the transmitted beam enters into the objective
at a given angle $\alpha$.  Thanks to the objective lens, a plane close to the
sample is imaged on the 2D sensor, and this ensures that both
the scattered and transmitted beams
are overlapping and collected on the sensor. 

\begin{figure}[h]
\begin{center}
\begin{tabular}{cc}
\includegraphics{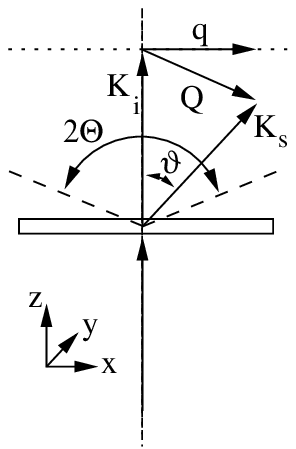}
&
\includegraphics{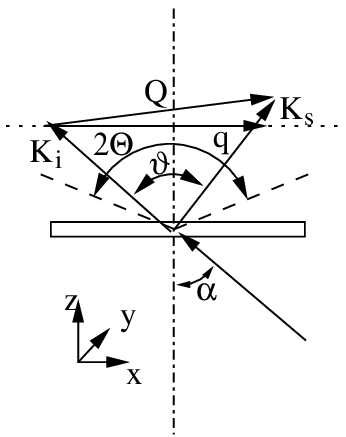}
\\
\includegraphics{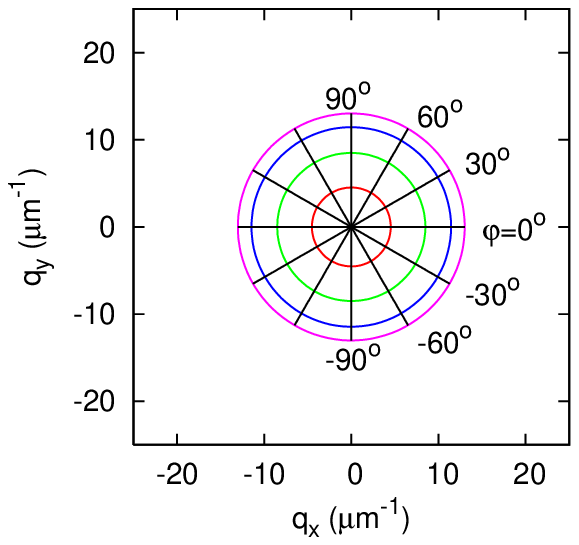}
&
\includegraphics{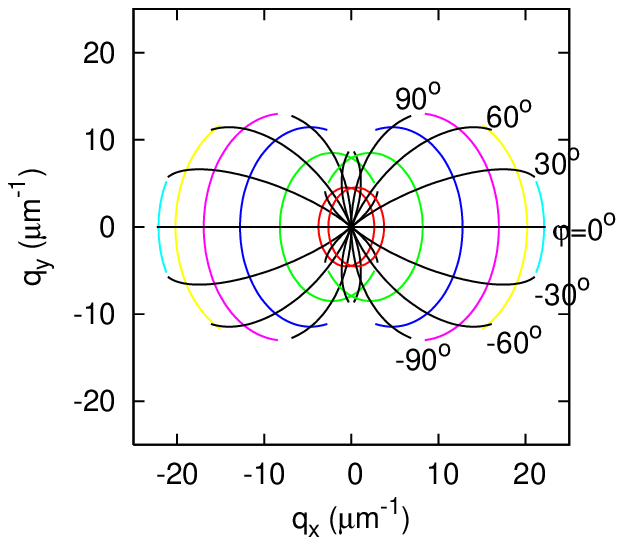}
\end{tabular}

\includegraphics{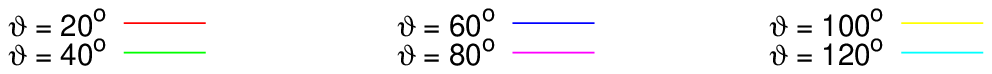}
\end{center}
\caption{Upper panels: schematic view of the geometry of the wave
vectors corresponding to the incoming light ($\vect{K}_i$), the scattered
light ($\vect{K}_s$), the transferred wave vector ($\vect{Q}$), and the 2D
image wave vector ($\vect{q}$). $2\Theta$ is the maximum acceptance angle 
of the objective and $\alpha$ is the tilted angle. 
Lower panels: mapping of the IPS wave vector
$\vect{q}$ as a function of scattering angle $\vartheta$ and azimuthal
angle $\varphi$. Acceptance angle $\Theta=80^{\circ}$.
Left panels: collinear illumination. Right
panels: out-of-axis illumination.} 
\label{fig_vettori}
\end{figure}

We now discuss how to relate the value of the image wave vector $\vect{q}$
with the transferred wave vector $\vect{Q}$,
which have to be taken into account when performing a TLM measurement.
Fig.~\ref{fig_vettori}, upper panels, shows a
schematic drawing of the geometrical arrangement of the wave
vectors, for the two cases of collinear (left panels) and out-of-axis
(right panels) illumination with tilt angle $\alpha$. 
The picture helps understand the
physical mechanism by which TLM increases the range of collected scattering
angles.

Consider an objective with a given numerical aperture, i.e. able to
collect beams within an angle $\Theta$ with respect to its optical
axis. If the illuminating beam is sent along the optical axis, then
the maximum detectable scattering angle $\vartheta$ is at most
$\Theta$. 

TLM allows to nearly double this range. Indeed, with this
geometry, the objective gathers the illuminated beam at an angle
close to $\Theta$, together with the light scattered at the opposite
side of the objective (see Fig.
\ref{fig_vettori}, upper row, right panel). 
In this case the angle
between the transmitted beam and the scattered one can be close to
$2\Theta$, taking advantage of the objective's total numerical
aperture.

A SINF measurement consists in collecting several images and then
calculating their IPS $S\left(\vartheta, \varphi\right)$, that is
the near field Image Power Spectrum, obtained, through a FFT algorithm, 
as the mean square of the Fourier transform of the images \cite{brogioli2002}.

Each point in the IPS $S\left(\vect{q}\right)$ represents the
amplitude of the Fourier mode with wave
vector $\vect{q}$ of the near field images; 
or equivalently, it represents the light intensity
$I\left(\vartheta, \varphi\right)$ at a given
scattering angle $\vartheta$ and azimuthal angle $\varphi$.

In the following, we will describe the relationship between
$\vect{q}$ and $\left(\vartheta, \varphi\right)$, which is essential to relate
the obtained IPS $S\left(\vect{q}\right)$ to the scattered intensity
$I\left(\vartheta, \varphi\right)$:
\begin{equation}
S\left[\vect{q}\left(\vartheta, \varphi\right)\right] = T\left(\vartheta, \varphi\right) I\left(\vartheta, \varphi\right)
\label{eq_IPS_inst}
\end{equation}
where $T\left(\vartheta, \varphi\right)$ is the tranfer function of the optical system.

Fig.~\ref{fig_vettori} shows the geometrical arrangement for the
scattering wave vector $\vect{K}_s$, the impinging beam wave vector
$\vect{K}_i$, the transferred wave vector
$\vect{Q}=\vect{K}_s-\vect{K}_i$ and the 2D image wave vector
$\vect{q}$. The image wave vector $\vect{q}$ is actually the
2D projection of the transferred wave vector $\vect{Q}$ on the plane
perpendicular to the optical axis, i.e. the sensor plane. 
Therefore the $x$ and $y$ components of the two vectors 
$\vect{Q}$ and $\vect{q}$ are the same.

If the transferred wave vector $\vect{Q}$ is known, then the wave
vector $\vect{q}$ on the image can be directly obtained:
\begin{equation}
\vect{q}\left(\vect{Q}\right) = \left(Q_x, Q_y, 0\right)
\end{equation}
Using this equation, $\vect{q}$ can also be expressed in terms of
the scattering angle $\vartheta$, azimuthal angle $\varphi$ and the tilt
angle $\alpha$ as:
\begin{equation}
\vect{q}\left(\vartheta,\varphi\right) = K \left[
\cos\left(\alpha\right) \sin\left(\vartheta\right) \cos\left(\varphi\right) +
\sin\left(\alpha\right) \cos\left(\vartheta\right) -
\sin\left(\alpha\right),
\sin\left(\vartheta\right) \sin\left(\varphi\right),
0
\right]
\end{equation}

The inverse problem is somewhat tricker. If the image wave
vector $\vect{q}$ is given, that is the actual experimental
situation, then a third equation is necessary to get the third
$z$ component of the transferred wave vector $\vect{Q}$. This can be
obtained from the condition
$\left|\vect{Q}+\vect{K}_i\right|=\left|\vect{K}_s\right|=K$, plus
the condition $\vect{K}_s \cdot \hat{z}>0$ which allows choosing one
of the two solutions of the quadratic equation. The relationship
between the image and the transferred vector is therefore:
\begin{equation}
\vect{Q}\left(\vect{q}\right) = \left(q_x, q_y,
\sqrt{\left(\vect{K}_i \cdot \hat{z}\right)^2-\vect{q}^2-2\vect{q}\cdot\vect{K}_i}
- \vect{K}_i \cdot \hat{z}
\right)
\label{eq_elegante}
\end{equation}

The graphs of $\vect{q}$ for values of scattering angle $\vartheta$
from $20^{\circ}$ up to $120^{\circ}$ and azimuthal angle $\varphi$
in steps of $30^{\circ}$ are shown in Fig.~\ref{fig_vettori}, lower
panels, for collinear and tilted illumination. They represent the
mapping of $\vartheta, \varphi$ on IPS images. It's worth noting
that each scattered beam at $\vartheta, \varphi$ corresponds to
\emph{two} points, $\vect{q}$ and $-\vect{q}$, in the IPS. The
mapping $-\vect{q}\left(\vartheta, \varphi\right)$ is also shown in
Fig.~\ref{fig_vettori}, represented by thin lines. It is evident that the
out-of-axis illumination allows mapping of angles larger than
$90^{\circ}$ in the IPS.

It's worth noting that very large scattering angles $\vartheta$ are
accessible only for azimuthal angles $\varphi$ nearly aligned with
the illumination tilt plane, that is $\varphi \approx 0^{\circ}$. In
the following, the mapping $\vect{q}\left(\vartheta, \varphi\right)$
will be used mainly to get the value of $I\left(\vartheta\right)$, that
is, the average value of $I\left(\vartheta\right)$ over a small
range of $\varphi$ around 0. If the polarization of the main beam is
$\varphi=0^{\circ}$, the measured scattering component will be
called ``parallel''. In the opposite case, that is
$\varphi=90^{\circ}$, we measure the ``perpendicular'' component.

\section{Data processing.}

By using Eq.~\ref{eq_elegante} it is possible to relate the output of the
Fourier analysis of the grabbed images $S\left(q\right)$
to the scattering intensity $I\left(Q\right)$. In order to evaluate the sample dynamics,
further analysis can be implemented by using the so called Exposure-Time-Dependent Spectrum (ETDS) processing,
which we recently reported in \cite{brogioli2008}. Essentially, the IPS is evaluated by taking
images at various exposure times, thus actually obtaining the
Exposure-Time-Dependent Spectrum $S\left(q,\Delta t\right)$
\cite{oh2004}. As $\Delta t$ is increased, the fluctuations
average out, leading to a decrease in the ETDS signal. For example,
for Brownian colloidal particles, the ETDS decreases strongly at large
scattering angles $\vartheta$. For very long $\Delta t$, all the
fluctuations are washed out, and the resulting images contain only the
optical background plus the fluctuating instrumental noise
$B\left(q\right)$. In this way, the ETDS gives access to a
quantitative measurement of the system dynamics. Indeed, it is possible to
analytically derive a theoretical expression for the $S\left(q,\Delta t\right)$, and thus
to get the diffusion coefficient of the
particles by using a fitting procedure \cite{oh2004,brogioli2008}. When a
negligible $\Delta t$ is considered, the ``instantaneous'' IPS
$S\left(q\right)$ is obtained, as described by Eq. \ref{eq_IPS_inst}.
When $\Delta t$ is not negligible with respect to
the sample dynamics, the ETDS 
is accordingly modified \cite{brogioli2008}. For colloidal samples, the time correlation
function is a decreasing exponential with decay time $\tau=1/(D
Q^2)$, where $D$ is the translational diffusion coefficient of the
nanoparticles. In this case the ETDS is expressed as \cite{brogioli2008,oh2004}:

\begin{equation}
\label{eq_ETD} S\left[q\left(\vartheta, \varphi\right),\Delta
t\right] = T\left(\vartheta, \varphi\right) I\left(\vartheta,
\varphi\right) f\left(\frac{\Delta t}{\tau}\right)
\end{equation}
where:
\begin{equation}
\label{eq_ETDf} f\left(x\right)=2\frac{e^{-x}-1+x}{x^2}
\end{equation}

The procedure for obtaining the decay time $\tau$ consists in
acquiring a series of images with many different exposure times, in
analyzing the ETDS $S\left[q\left(\vartheta, \varphi\right),\Delta
t\right]$ for each scattering angle as a function of the exposure
time, and finally in fitting it with Eq. \ref{eq_ETD}, keeping $\tau$ as free
parameter. By using this procedure, a direct measurement
of the time constants for all the measured wave
vectors is simultaneously achieved. Moreover, the static scattered intensity
$I\left(\vartheta, \varphi\right)$, multiplied by the transfer
function $T\left(\vartheta, \varphi\right)$ is obtained, in analogy
with the procedure developed for the calculation of the structure
function in \cite{croccolo2006,croccolo2006bis}. In
general, the instrument transfer function $T\left(\vartheta,
\varphi\right)$ can significantly depend on the wave vector, such as in
the shadowgraph \cite{croccolo2007}. On the contrary, in the near
field scattering regime, the transfer function is slowly varying
within the accessible scattering range \cite{ferri2004}. The
transfer function was actually
evaluated by performing calibration measurements on known
samples, so that a direct access to the static scattered intensity
of our sample can be obtained.

\section{Experimental set up.}

\begin{figure}[h]
\begin{center}
\includegraphics{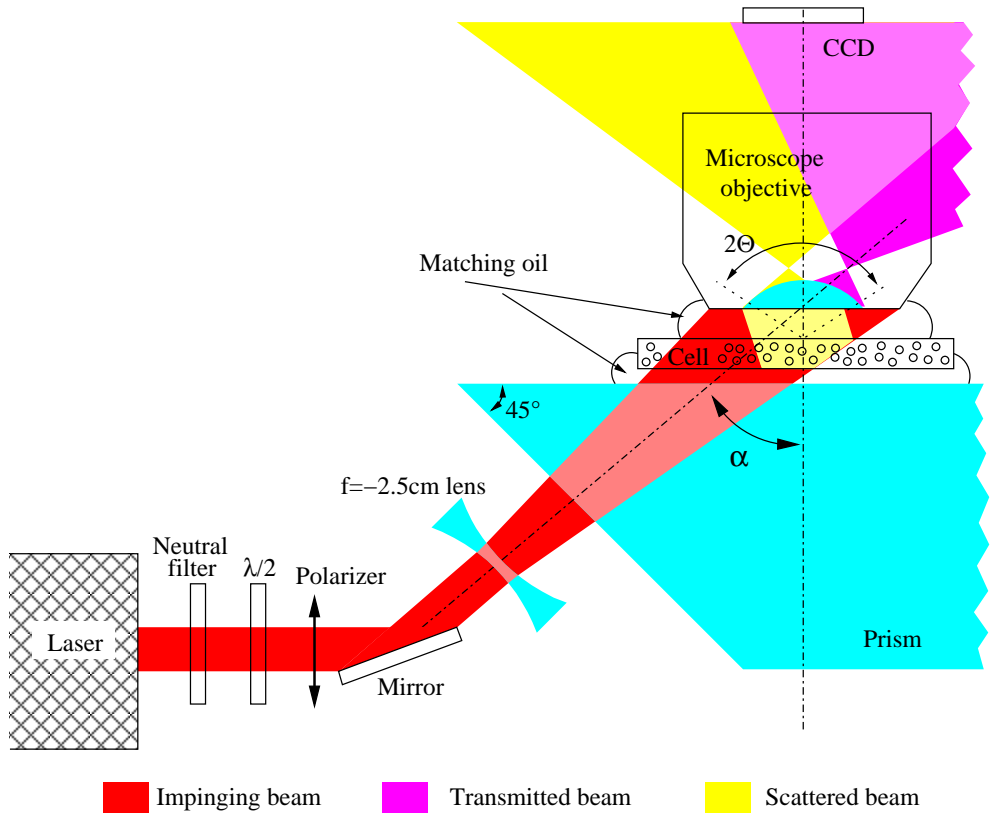}
\end{center}
\caption{Sketch of the optical set-up. The He-Ne laser generates a
collimated laser beam, which is attenuated by a neutral filter. A
half-wave plate and a polarizer control the beam polarization. A
mirror bends and adjusts the beam direction. The beam is then
expanded by means of a negative focal-length lens, making it
slightly divergent. The beam goes through a $45^{\circ}$ prism, 
and enters the sample cell with about a $\alpha=45^{\circ}$ angle with
respect to the vertical. Scattered light is acquired in the near
field, together with transmitted light, through a vertical
microscope objective, which conjugates a plane close to the sample
onto the CCD sensor.} 
\label{fig_setup}
\end{figure}

In Fig.~\ref{fig_setup} a sketch of the optical set-up is
shown. The experimental apparatus consists of a standard microscope
equipped with a low-speed CCD camera (Andor Luca). The microscope
core is made by a commercial infinite conjugate oil immersion
objective (Nikon plan achromat) 100X, numerical aperture NA=1.25 and
working distance 0.17 mm, used with a tube lens with half the
nominal focal, so that the resulting magnification is 50X. Its
acceptance angle is $\Theta=70^{\circ}$. Some figures in the present
article report images and data obtained with a plan-achromatic 40X
objective (Optika Microscopes, FLUOR), with 0.65 numerical aperture,
to emphasize the effects of the finite acceptance angle
$\Theta=30^{\circ}$.

The illuminating source of the microscope has been replaced by a
$10\mathrm{mW}$ He-Ne laser (Nec), enlarged by a negative $-2.5
\mathrm{cm}$ focal-length lens, making the beam diverge, and thus
increasing its diameter up to about $4\mathrm{mm}$ at the sample
plane. This use of a single lens instead of a beam expander introduces
a small error of about $1^{\circ}$, which is negligible with respect to
the large scattering angles we are interested in.
The beam coming from the laser source is attenuated by a
variable neutral-filter wheel, with transmission range $0.3 -
0.0003$, and is bent by a mirror at a fixed angle respect to the
vertical axis. The polarization state of the impinging beam is
controlled by a half-wavelength plate and a polarizer, placed
perpendicularly to the beam and suitably oriented. After
$20\mathrm{cm}$ the diverging beam enters a $45^{\circ}$ prism, in
contact with the cell. The beam direction inside the cell is
$\alpha=45^{\circ}$ with respect to the vertical. Without the aid of
such a prism the maximum tilting angle is bound by the total
reflection at the glass/air interface.

The scattered light is collected beyond the sample, together with
the transmitted beam (heterodyne detection) in the near field by
means of the microscope objective and the tube lens conjugating a
plane close to the sample onto the CCD sensor plane.

The sample is placed in a glass cell of $1\mathrm{mm}$ optical path,
made by two microscope cover slips, spaced by small glass strips cut
from microscope slides, and glued with silicone rubber. Both
couplings (prism/cell and cell/objective) are index matched by a
standard microscope oil, with refractive index 1.55. The first focal
plane of the objective is placed inside the sample, the exact
position being not an important issue. The large beam size ensures
that the imaged area receives light from a portion of illuminated
sample, for any angle inside the range accepted by the objective,
thus also satisfying the near field condition \cite{magatti2008}.

\section{Materials}
We studied different samples of commercial
polystyrene calibrated spherical nano-particles. The different samples are
listed in Tab.~\ref{tab_samples}.
\begin{table}
\begin{tabular}{c|l|c|c}
Sample & Producer & Nominal diameter (nm) & Dynamic SIFF measured diameter (nm) \\
\hline
A & Polyscience Inc. & 20 & 23 \\
B & Duke Scientific Corp. & 80 & 81 \\
C & Duke Scientific Corp. & 150 & 149 \\
D & Polyscience Inc. & 400 & 402 \\
E & Polyscience Inc. & 1520 & 1620
\end{tabular}
\caption{List of the analyzed samples.}
\label{tab_samples}
\end{table}
The particles are dispersed in deionized water,
and are used without further purification, neither dialysis nor salt adding.
The particles are strictly mono-dispersed 
and their nominal sizes have been checked by
a traditional dynamic SIFF apparatus, providing measured
sizes shown in fourth column Tab.~\ref{tab_samples}.
Samples B and C, from Duke Scientific Corporation, are dynamic light scattering
calibration standards;  
the measured diameters are in excellent agreement with nominal values, and the 
error is compatible with the variability between different samples, as reported
by the producer. The same excellent agreement is found also for sample D
from Polyscience. For what concerns samples A and E the error is larger, of the order of 10\%,
probably due to the aging. From now on, we will compare the data obtained with the TLM apparatus
with the above data actually measured with dynamic SIFF, and not with the nominal values.

\section{Results and discussion.}

\begin{figure}[h]
\begin{center}
\begin{tabular}{cccc}
$\alpha = 0^{\circ}$ & $\alpha = 12^{\circ}$ & $\alpha = 24^{\circ}$ & $\alpha = 45^{\circ}$
\\
\includegraphics[scale=0.4]{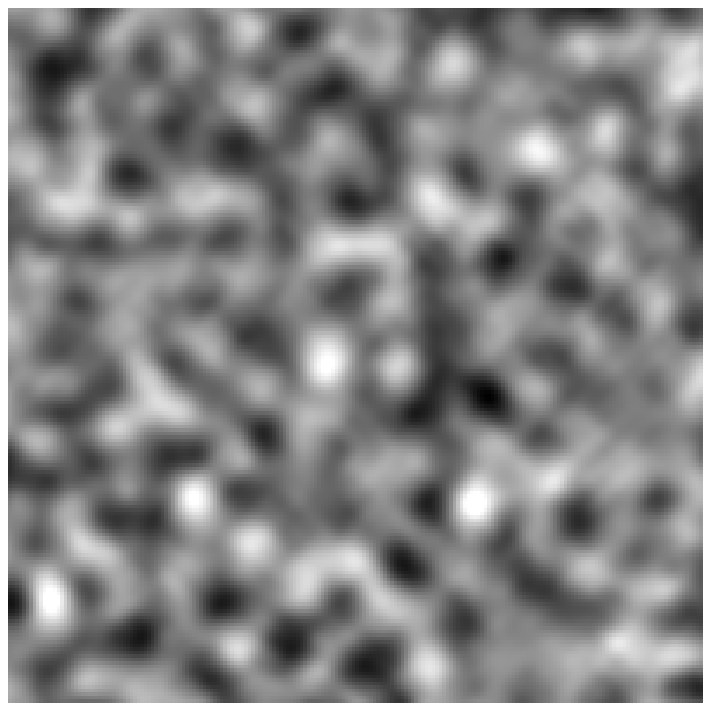}&
\includegraphics[scale=0.4]{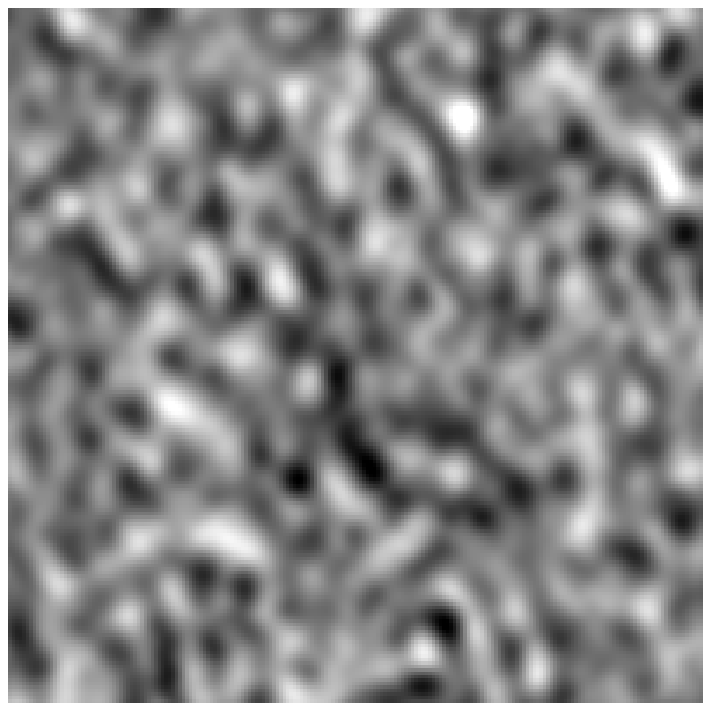}&
\includegraphics[scale=0.4]{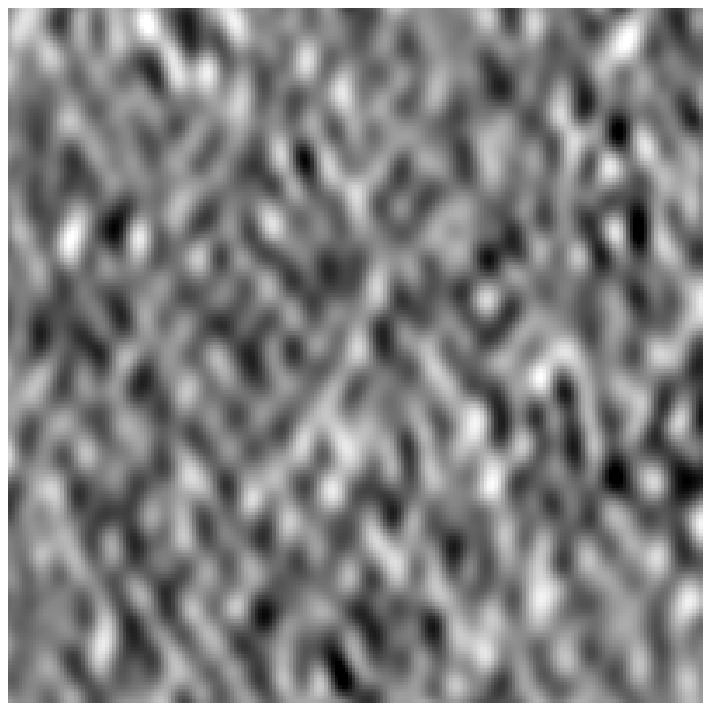}&
\includegraphics[scale=0.4]{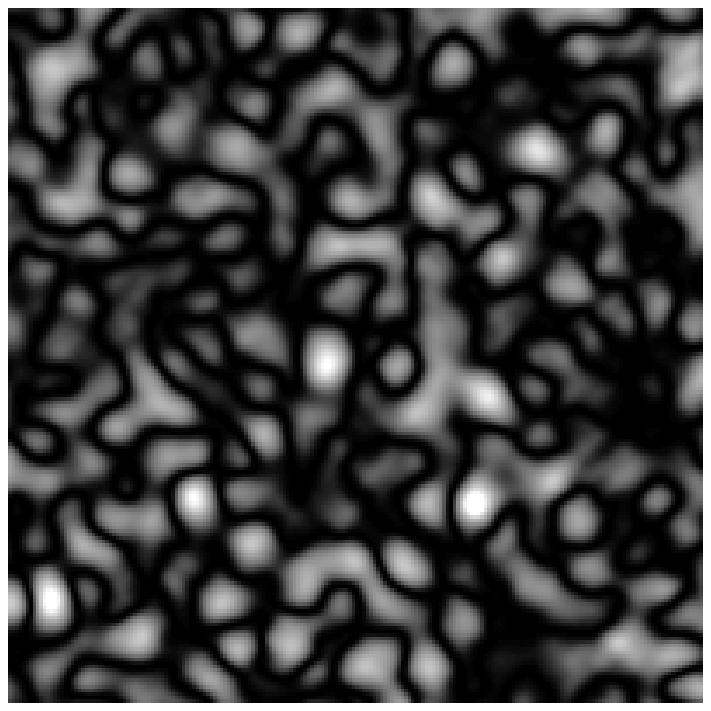}
\\
\includegraphics[scale=0.4]{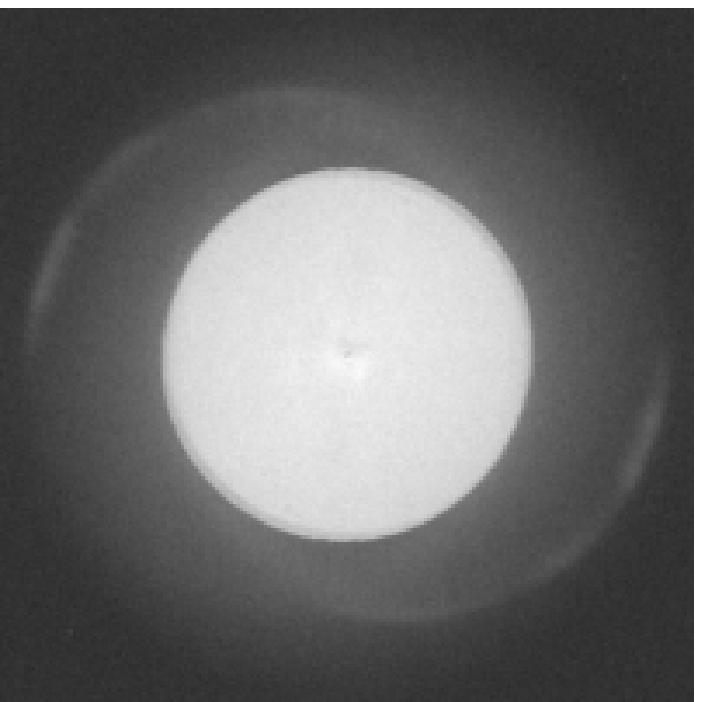}&
\includegraphics[scale=0.4]{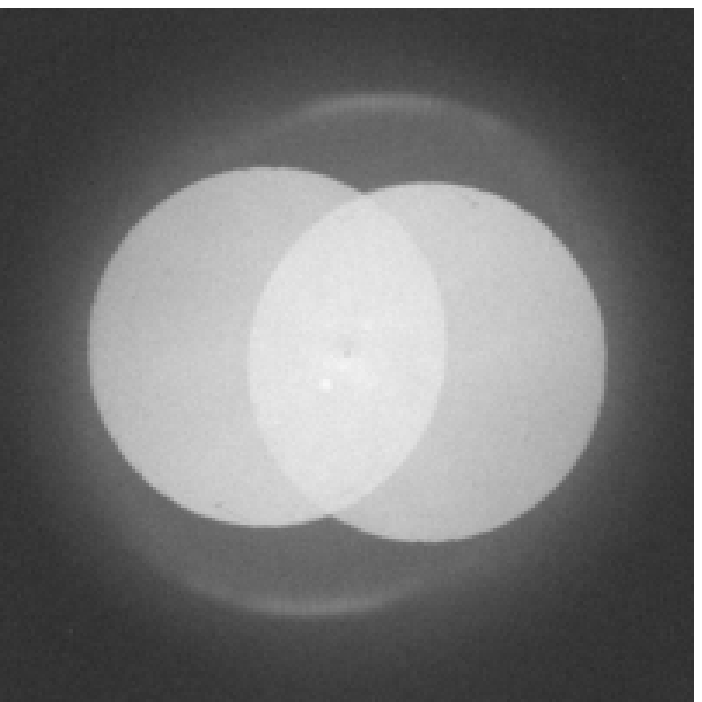}&
\includegraphics[scale=0.4]{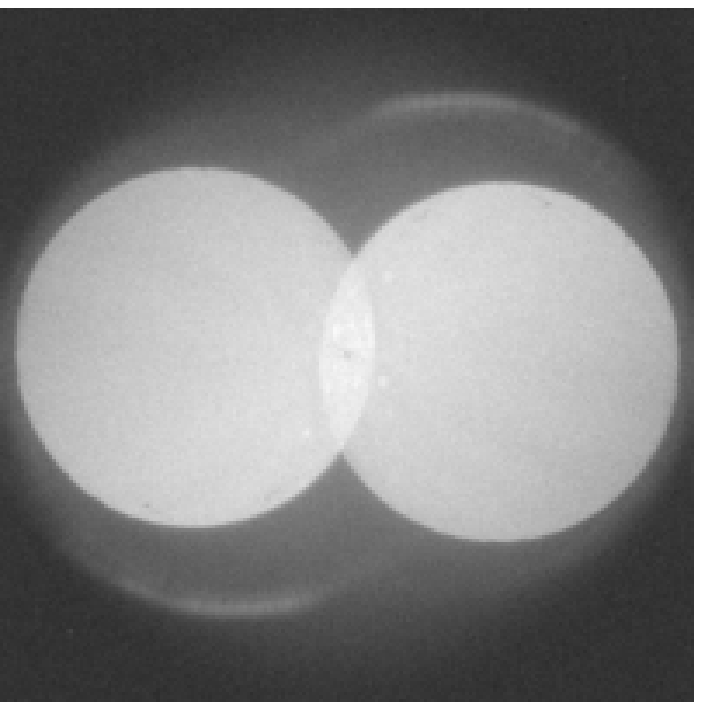}&
\includegraphics[scale=0.4]{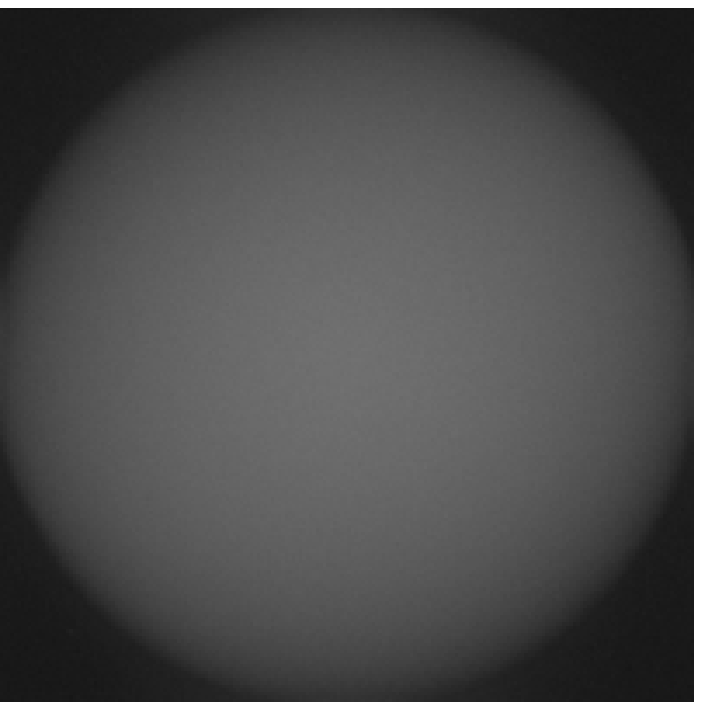}
\end{tabular}
\end{center}
\caption{Upper panels: 2D images of the SINF speckle field, taken at different
tilt angles $\alpha$. The size of the images is $12.5\mathrm{\mu m}$ in real space.
After software magnification, finite-pixel-size effects have
been removed by suitable image processing for an easier visualization.
The contrast of the image with $\alpha=45^{\circ}$ has been enhanced.
Lower panels: power spectra of the same images, 
presented using a logarithmic intensity scale.
The size of the image represents a wave vector $q=12\mathrm{\mu m}^{-1}$.
The microscope objective is a 40X, 0.65 NA. 
The bright disks represent the heterodyne signal, given by the interference
between the most intense transmitted beam and 
the scattered beam. The geometrical position
of the disks depends on the tilt angle $\alpha$.
A faint circular halo, centered in the IPS, represents the
negligible homodyne signal, due to self 
interference between different scattered beams.
The disks are surrounded
by very faint whiskers due to spurious reflections.
In the last column on the right, the image 
shows the homodyne contribution only,
since the transmitted beam is
stopped by the objective diaphragm, by deliberately
using a tilting angle higher than the objective maximum acceptance angle.
This last configuration cannot
be used in TLM setup.
Data obtained with sample C.
}
\label{fig_omodina}
\end{figure}
In Fig.~\ref{fig_omodina}, upper panels, typical near field images 
at different tilt angles $\alpha$ are shown as collected by a CCD, while
in the lower panels the corresponding power spectra (IPS) are
presented. We used for these images the sample C, which provides a
sufficiently intense scattering almost on
every scattering angle $\vartheta$. The employed objective is 40X, with an
acceptance angle $\Theta=30^{\circ}$.

For collinear illumination ($\alpha=0$), the IPS shows a bright,
nearly uniform circle at the center, corresponding to the nearly
isotropically scattered light, clipped by the angular acceptance
of the objective. The beams scattered at small angle $\vartheta$ are
represented by the points close to the center of the IPS image. 
In general, the optical collecting system determines 
the scattering angle range which can be measured.
In SIFF, this angular limitation produces a vignetting visible in the far
field image. By comparison, as shown in Fig.~\ref{fig_omodina}, 
in SINF set up the acceptance angle of the microscope
generates a vignetting in the near field IPS (Fourier space).
On the other hand, the intensity distribution in the near field
image (real space) shows a speckle pattern, superimposed on an
average uniform intensity distribution with no intensity fading
in the periphery of the image.

For out-of-axis illumination, the center of the IPS still represents
beams with small scattering angle $\vartheta$, 
but, as the tilt angle $\alpha$ is increased, the vignetting
circle is translated with respect to the center of the IPS, 
as previously shown in Fig.~\ref{fig_vettori}. 
This allows collecting light at larger scattering angles. It's worth
pointing out that the IPS is always center-symmetric, so that
\emph{two} clipping circles are visible in case of out-of-axis
illumination. In this situation the centers of the shifted circles 
represent the beam
scattered along the optical axis of the objective.

In Fig.~\ref{fig_omodina}, 
for every in-axis and out-of-axis images,
a faded circular halo outside the brighter internal
circles is also visible, due to the self interference
(homodyne term) of the scattered light collected by the objective.
Obviously, in our heterodyne detection scheme,
the homodyne signal is not the object of our study
and its intensity is negligible in the shown cases.
The radius of the homodyne signal disk is twice the radius of
the heterodyne disk. As apparent from
Fig.~\ref{fig_omodina}, the size and position of the homodyne
scattering circle is unaffected by the tilt angle $\alpha$.

In the following, we quantitatively explain the observed characteristics
of the homodyne and heterodyne disks.
The objective collects beams inside its aperture cone,
that is, beams whose wave vector $\vect{K}_s=\left(K_x, K_y, K_z\right)$ 
satisfies the condition:
\begin{equation}
\sqrt{K_x^2 + K_y^2}<K \sin \Theta.
\label{eq_cono}
\end{equation}
When considering a transmitted beam with wave vector 
$\vect{K}_i=K\left(\sin \alpha, 0, \cos \alpha\right)$,
the transferred wave vector $\vect{Q}=\vect{K}_s-\vect{K}_i$ satisfies:
\begin{equation}
\sqrt{\left(Q_x + K \sin \alpha\right)^2 + Q_y^2}<K \sin \Theta,
\end{equation}
which describes the heterodyne signal disk, centered in 
$\vect{q}=\left(K \sin \alpha,0\right)$ with radius $K \sin \Theta$.

When considering the interference between a couple of scattered 
beams, with wave vectors $\vect{K}_s=\left(K_x, K_y, K_z\right)$ 
and $\vect{K}_s'=\left(K_x', K_y', K_z'\right)$, they both satisfy 
Eq. (\ref{eq_cono}). The interference pattern is again a Fourier
mode with wave vector $\vect{q}=\left(K_x-K_x', K_y - K_y', 0\right)$. 
In this case, the condition on $\vect{q}$ becomes:
\begin{equation}
\sqrt{Q_x^2 + Q_y^2}<2 K \sin \Theta,
\end{equation}
which represents the homodyne signal disk, centered in 
$\vect{q}=\left(0, 0\right)$ and with radius $2 K \sin \Theta$,
twice than in the heterodyne case.

The right column in Fig.~\ref{fig_omodina} refers to a tilt angle
$\alpha$ larger then the angular acceptance $\Theta$, which cannot be used
for TLM detection. In this case the
transmitted beam is no longer collected by the objective, and the
heterodyne signal disappears. Correspondingly, the near field image
shows the well known characteristics of the homodyne speckles, appearing
as bright spots on a dark background. By contrast, the heterodyne speckles 
shown in the first three upper panels of Fig.~\ref{fig_omodina} appear,
as usual, as an even mixing of bright and dark spots superimposed on an average intensity.
This distinction between 
heterodyne and homodyne speckles could be fully appreciated by considering 
the histogram of the intensity distribution of the speckles,
showing an exponential decay for the homodyne speckles, and a gaussian-like
for the heterodyne speckles (data not shown).

\begin{figure}[h]
\begin{tabular}{c}
\begin{tabular}{cc}
40X objective&100X objective, 1/2X tube lens\\
\setlength{\unitlength}{0.120450pt}
\begin{picture}(1484,1259)(0,0)
\put(560,480){\includegraphics[width=56.1297pt]{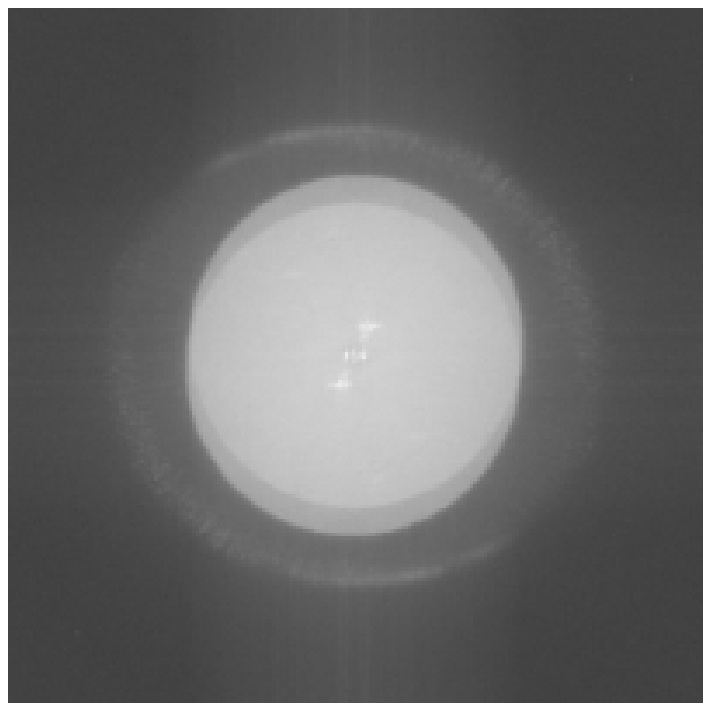}}\footnotesize
\color{black}
\thicklines \path(329,342)(370,342)
\thicklines \path(1258,342)(1217,342)
\put(288,342){\makebox(0,0)[r]{-20}}
\color{black}
\thicklines \path(329,528)(370,528)
\thicklines \path(1258,528)(1217,528)
\put(288,528){\makebox(0,0)[r]{-10}}
\color{black}
\thicklines \path(329,713)(370,713)
\thicklines \path(1258,713)(1217,713)
\put(288,713){\makebox(0,0)[r]{ 0}}
\color{black}
\thicklines \path(329,899)(370,899)
\thicklines \path(1258,899)(1217,899)
\put(288,899){\makebox(0,0)[r]{ 10}}
\color{black}
\thicklines \path(329,1085)(370,1085)
\thicklines \path(1258,1085)(1217,1085)
\put(288,1085){\makebox(0,0)[r]{ 20}}
\color{black}
\thicklines \path(422,249)(422,290)
\thicklines \path(422,1178)(422,1137)
\put(422,166){\makebox(0,0){-20}}
\color{black}
\thicklines \path(608,249)(608,290)
\thicklines \path(608,1178)(608,1137)
\put(608,166){\makebox(0,0){-10}}
\color{black}
\thicklines \path(793,249)(793,290)
\thicklines \path(793,1178)(793,1137)
\put(793,166){\makebox(0,0){ 0}}
\color{black}
\thicklines \path(979,249)(979,290)
\thicklines \path(979,1178)(979,1137)
\put(979,166){\makebox(0,0){ 10}}
\color{black}
\thicklines \path(1165,249)(1165,290)
\thicklines \path(1165,1178)(1165,1137)
\put(1165,166){\makebox(0,0){ 20}}
\color{black}
\color{black}
\thicklines \path(329,249)(1258,249)(1258,1178)(329,1178)(329,249)
\color{black}
\put(82,713){\makebox(0,0)[l]{\rotatebox[origin=c]{90}{$q_y (\mathrm{\mu m}^{-1})$}}}
\color{black}
\put(793,42){\makebox(0,0){$q_x (\mathrm{\mu m}^{-1})$}}
\color{red}
\thinlines \path(793,630)(793,630)(799,630)(804,631)(809,632)(815,633)(820,634)(825,636)(830,638)(834,641)(839,643)(843,646)(848,650)(851,653)(855,657)(859,661)(862,665)(865,670)(868,674)(870,679)(872,684)(874,689)(875,694)(876,699)(877,705)(877,710)(878,715)(877,721)(877,726)(876,731)(874,736)(873,742)(871,746)(869,751)(866,756)(863,761)(860,765)(857,769)(853,773)(850,777)(845,780)(841,783)(837,786)(832,789)(827,791)(822,793)(817,795)(812,796)(807,797)(801,798)(796,798)
\thinlines \path(796,798)(791,798)(786,798)(780,797)(775,796)(770,795)(765,793)(760,791)(755,789)(750,786)(746,783)(742,780)(737,777)(734,773)(730,769)(727,765)(724,761)(721,756)(718,751)(716,746)(714,742)(713,736)(711,731)(710,726)(710,721)(709,715)(710,710)(710,705)(711,699)(712,694)(713,689)(715,684)(717,679)(719,674)(722,670)(725,665)(728,661)(732,657)(736,653)(739,650)(744,646)(748,643)(753,641)(757,638)(762,636)(767,634)(772,633)(778,632)(783,631)(788,630)(793,630)
\color{green}
\thinlines \path(793,558)(793,558)(804,558)(813,559)(823,561)(833,563)(843,566)(852,569)(861,573)(870,578)(879,583)(887,589)(895,595)(902,601)(910,609)(916,616)(922,624)(928,632)(933,641)(937,650)(941,659)(944,669)(947,678)(949,688)(950,698)(951,708)(951,718)(951,728)(950,738)(948,748)(946,758)(943,767)(939,777)(935,786)(930,795)(925,803)(919,811)(913,819)(906,826)(899,833)(891,840)(883,846)(875,851)(866,856)(857,860)(848,864)(838,867)(828,870)(818,871)(809,873)(799,873)
\thinlines \path(799,873)(788,873)(778,873)(769,871)(759,870)(749,867)(739,864)(730,860)(721,856)(712,851)(704,846)(696,840)(688,833)(681,826)(674,819)(668,811)(662,803)(657,795)(652,786)(648,777)(644,767)(641,758)(639,748)(637,738)(636,728)(636,718)(636,708)(637,698)(638,688)(640,678)(643,669)(646,659)(650,650)(654,641)(659,632)(665,624)(671,616)(677,609)(685,601)(692,595)(700,589)(708,583)(717,578)(726,573)(735,569)(744,566)(754,563)(764,561)(774,559)(783,558)(793,558)
\color{blue}
\thinlines \path(793,505)(793,505)(807,506)(820,507)(834,509)(847,512)(860,516)(873,521)(885,526)(897,532)(909,539)(920,547)(930,555)(940,564)(950,574)(959,584)(967,595)(974,606)(981,618)(987,630)(992,642)(997,655)(1000,668)(1003,681)(1005,694)(1006,708)(1006,721)(1006,735)(1004,748)(1002,762)(999,775)(995,788)(990,800)(984,812)(978,824)(971,836)(963,847)(954,857)(945,867)(935,877)(925,885)(914,893)(903,900)(891,907)(879,913)(866,918)(853,922)(840,925)(827,928)(814,930)(800,931)
\thinlines \path(800,931)(787,931)(773,930)(760,928)(747,925)(734,922)(721,918)(708,913)(696,907)(684,900)(673,893)(662,885)(652,877)(642,867)(633,857)(624,847)(616,836)(609,824)(603,812)(597,800)(592,788)(588,775)(585,762)(583,748)(581,735)(581,721)(581,708)(582,694)(584,681)(587,668)(590,655)(595,642)(600,630)(606,618)(613,606)(620,595)(628,584)(637,574)(647,564)(657,555)(667,547)(678,539)(690,532)(702,526)(714,521)(727,516)(740,512)(753,509)(767,507)(780,506)(793,505)
\color{black}
\thicklines \path(329,249)(1258,249)(1258,1178)(329,1178)(329,249)
\end{picture}
&
\setlength{\unitlength}{0.120450pt}
\begin{picture}(1484,1259)(0,0)
\put(501,421){\includegraphics[width=70.3428pt]{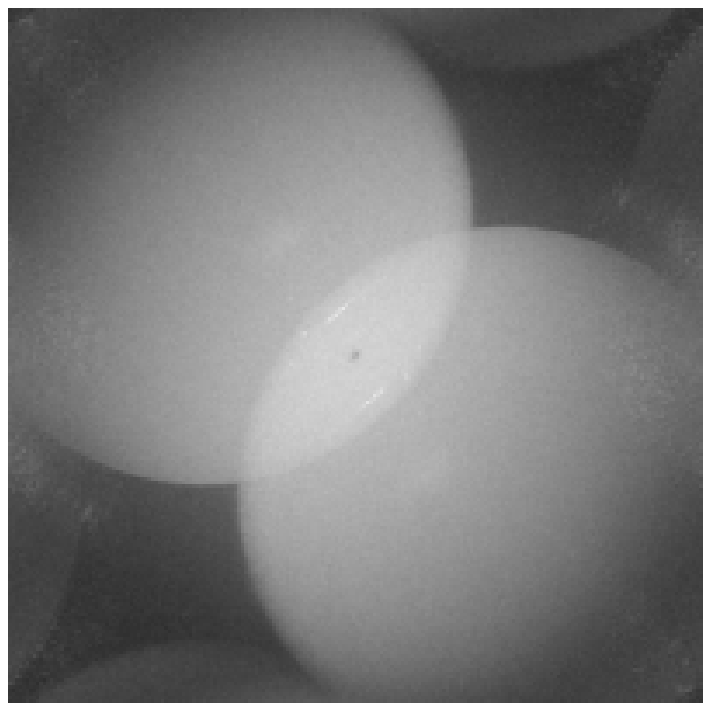}}\footnotesize
\color{black}
\thicklines \path(329,342)(370,342)
\thicklines \path(1258,342)(1217,342)
\put(288,342){\makebox(0,0)[r]{-20}}
\color{black}
\thicklines \path(329,528)(370,528)
\thicklines \path(1258,528)(1217,528)
\put(288,528){\makebox(0,0)[r]{-10}}
\color{black}
\thicklines \path(329,713)(370,713)
\thicklines \path(1258,713)(1217,713)
\put(288,713){\makebox(0,0)[r]{ 0}}
\color{black}
\thicklines \path(329,899)(370,899)
\thicklines \path(1258,899)(1217,899)
\put(288,899){\makebox(0,0)[r]{ 10}}
\color{black}
\thicklines \path(329,1085)(370,1085)
\thicklines \path(1258,1085)(1217,1085)
\put(288,1085){\makebox(0,0)[r]{ 20}}
\color{black}
\thicklines \path(422,249)(422,290)
\thicklines \path(422,1178)(422,1137)
\put(422,166){\makebox(0,0){-20}}
\color{black}
\thicklines \path(608,249)(608,290)
\thicklines \path(608,1178)(608,1137)
\put(608,166){\makebox(0,0){-10}}
\color{black}
\thicklines \path(793,249)(793,290)
\thicklines \path(793,1178)(793,1137)
\put(793,166){\makebox(0,0){ 0}}
\color{black}
\thicklines \path(979,249)(979,290)
\thicklines \path(979,1178)(979,1137)
\put(979,166){\makebox(0,0){ 10}}
\color{black}
\thicklines \path(1165,249)(1165,290)
\thicklines \path(1165,1178)(1165,1137)
\put(1165,166){\makebox(0,0){ 20}}
\color{black}
\color{black}
\thicklines \path(329,249)(1258,249)(1258,1178)(329,1178)(329,249)
\color{black}
\put(82,713){\makebox(0,0)[l]{\rotatebox[origin=c]{90}{$q_y (\mathrm{\mu m}^{-1})$}}}
\color{black}
\put(793,42){\makebox(0,0){$q_x (\mathrm{\mu m}^{-1})$}}
\color{red}
\thinlines \path(744,645)(744,645)(747,642)(750,640)(753,638)(757,637)(761,635)(765,634)(769,634)(773,633)(777,633)(781,633)(786,634)(790,634)(795,635)(799,637)(804,638)(808,640)(813,642)(817,645)(822,647)(826,650)(830,653)(834,656)(838,660)(842,664)(846,668)(849,672)(853,676)(856,680)(859,684)(861,689)(864,694)(866,698)(868,703)(869,707)(871,712)(872,717)(872,721)(873,726)(873,730)(873,735)(873,739)(872,743)(871,747)(870,751)(868,754)(866,758)(864,761)(862,764)(860,767)
\color{green}
\thinlines \path(845,565)(845,565)(854,569)(862,574)(870,580)(878,586)(886,592)(893,598)(900,605)(907,613)(914,620)(920,628)(926,636)(931,644)(936,653)(941,661)(945,670)
\color{blue}
\thinlines \path(920,511)(920,511)(931,519)(941,527)(951,536)(961,546)(970,556)(979,566)(987,576)(995,587)(1003,598)
\color{magenta}
\thinlines \path(982,462)(982,462)(994,471)(1006,481)(1017,492)(1027,503)(1037,515)(1047,527)(1056,539)
\color{yellow}
\thinlines \path(1038,429)(1038,429)(1050,440)(1061,450)(1071,461)(1081,473)(1091,485)
\color{black}
\thicklines \path(329,249)(1258,249)(1258,1178)(329,1178)(329,249)
\end{picture}
\end{tabular}
\\
\includegraphics{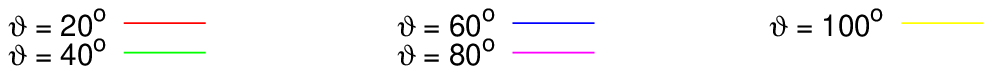}
\\
\begin{tabular}{cc}
\includegraphics{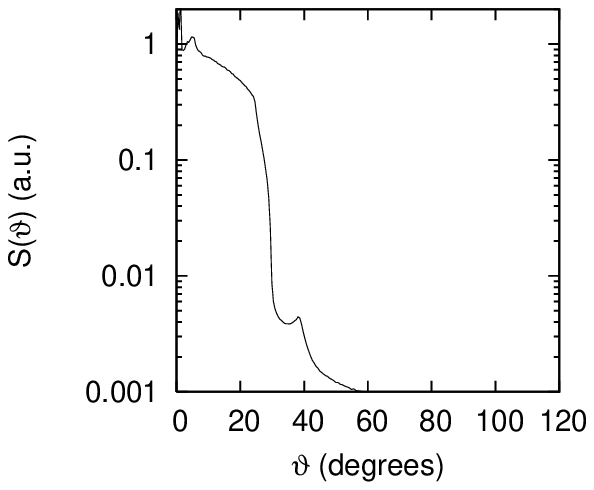}
&
\includegraphics{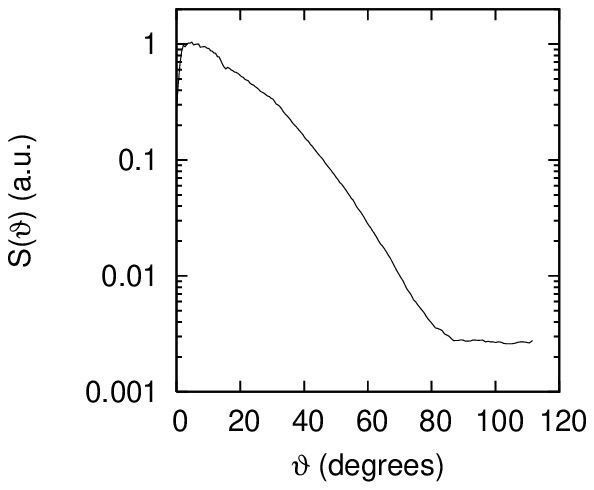}
\end{tabular}
\end{tabular}
\caption{
Upper panels: 2D representation of the Intensity Power Spectra IPS. 
Lower panels: spectrum intensity $S\left(\vartheta\right)$ as obtained by
averaging on the lines having the same $\vartheta$.
Left column: in-axis configuration, 40X objective, $\Theta = 30^{\circ}$. Right column: tilted configuration, 100X objective,
$\Theta = 70^{\circ}$, $\alpha=45^{\circ}$ with a tube lens with half the nominal length.
In the in-axis case, the power spectrum shows a dramatic drop around the angle $30^{\circ}$,
corresponding to the clipping due to the acceptance angle $\Theta=30^{\circ}$ of the 40X objective. On the contrary, the
100X objective, with tilted illumination, is able to collect light up to $110^{\circ}$.
Colored lines: lines of constant $\vartheta$ as in Fig.~\ref{fig_vettori}.
Data obtained with sample D.
}
\label{fig_scattering_range}
\end{figure}
Fig.~\ref{fig_scattering_range}, upper panels, shows the resulting
IPS obtained in two different configurations. In
Fig.~\ref{fig_scattering_range}, left column, the laser is collinear with the
objective's optical axis, $\alpha = 0$. In
Fig.~\ref{fig_scattering_range}, right column, the laser enters the sample tilted
of about $\alpha = 45^{\circ}$. In this second case, the CCD sensor
is also rotated by $45^{\circ}$ around the axis, perpendicularly to
the sensor plane, so that it is possible to achieve the largest
scattering vectors up to $110^{\circ}$ along the IPS diagonal. In
Fig.~\ref{fig_scattering_range}, lower panels, we show the IPS as
obtained by averaging over the colored lines which correspond to
to data with the same $\vartheta$ or with the same $\left|\vect{q}\right|$.
Actually the data are also averaged on
the region of small $\varphi$ in order to avoid mixing of the perpendicular
and parallel scattering contributions and to achieve larger scattering 
angle $\vartheta$.

In order to verify the validity of the proposed approach, 
we have performed several experiments on the samples listed in Tab.~\ref{tab_samples}.
For each sample seven series of $N=100$ images have been taken at different
exposure times $\Delta t$ ($0.55\mathrm{ms}$, $1.8\mathrm{ms}$,
$5.5\mathrm{ms}$, $18\mathrm{ms}$, $55\mathrm{ms}$,
$180\mathrm{ms}$, and $550\mathrm{ms}$), and we evaluated the 
static and dynamic power spectra of the sample.

\begin{figure}[h]
\begin{tabular}{cc}
\includegraphics{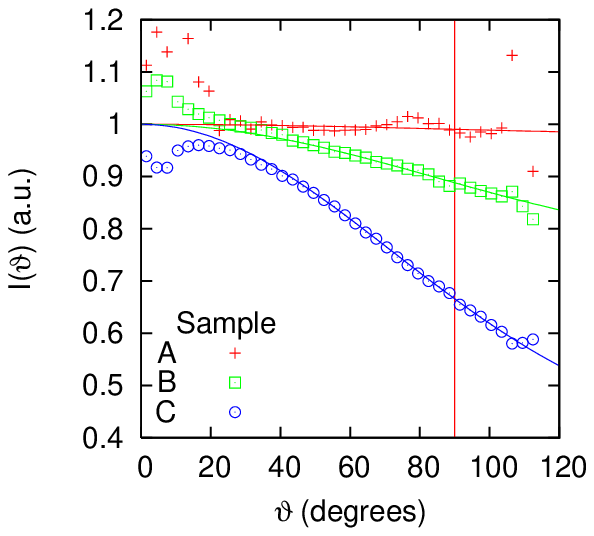}
&
\includegraphics{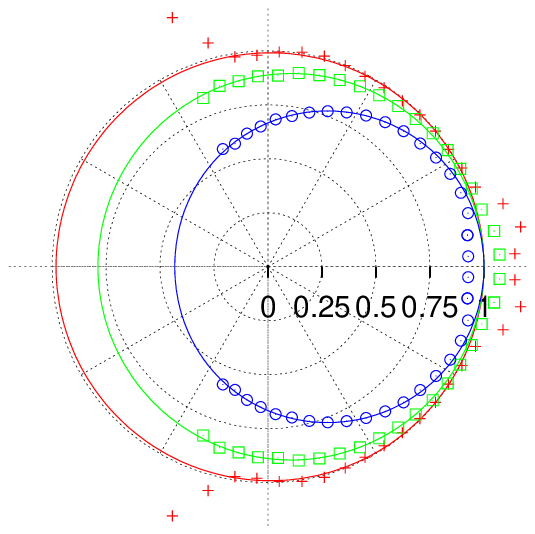}
\\
\includegraphics{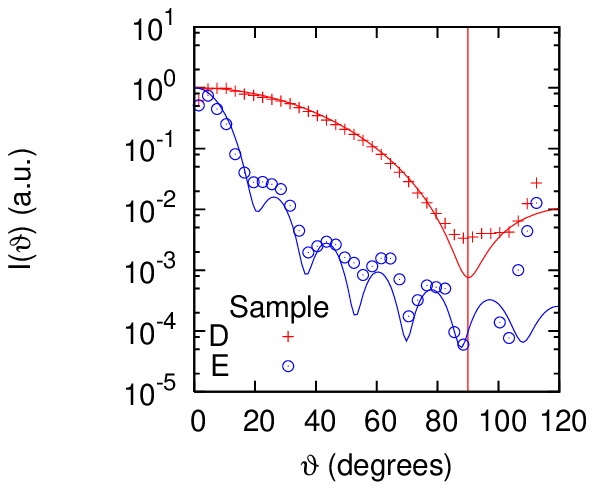}
&
\includegraphics{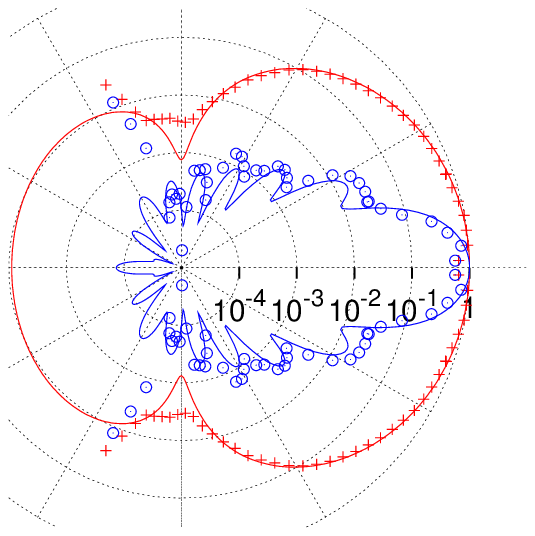}
\end{tabular}
\caption{
Scattered intensity $I\left(\vartheta\right)$ measured as a function of the scattering
angle $\vartheta$ in the plane perpendicular to the impinging polarization.
Experimental results for polystyrene colloids A, B, C, D, E.
The lines represent the values calculated with Mie theory for diameter $23\mathrm{nm}$,
$81\mathrm{nm}$, $149\mathrm{nm}$, $450\mathrm{nm}$, and $1700\mathrm{nm}$ respectively.
Left column graphs are in Cartesian coordinates, and right column graphs are in polar coordinates.
Upper panels represent scattering intensity in linear scale, while in the lower panels 
the scale is logarithmic.}
\label{fig_statico}
\end{figure}

In Fig.~\ref{fig_statico} we show the plots of the perpendicular scattering component 
$I(\vartheta)$ for different nanoparticles. The
instrument transfer function $T\left(\vartheta\right)$ has been
determined by measuring a reference sample, namely the sample A,
averaging 100 power spectra and smoothing the resulting $S\left(\vartheta\right)$.
Continuous lines illustrate the exact Mie theoretical results
\cite{Scott_web}. For samples A, B, C (Fig.~\ref{fig_statico}, upper panels)
we compare the experimental data with the Mie theoretical values by using
the diameters measured with dynamic SIFF and we obtain a reasonable agreement.
For samples D and E the value of the particle diameter 
necessary for nicely fit the data is larger than the value 
obtained by dynamic SIFF, by an amount of about 12.5\% and 5\% respectively.
The reason for this discrepancy is related to the difficulty inherent
to low intensity measurements spanning more than two orders of magnitude.
The vertical red line indicates the $90^{\circ}$ limit. The
similarity with Mie theory confirms the ability of the
proposed technique to measure the scattering signal up to
angles of $90^{\circ}$ and beyond. We note that, in the present work, we are
mainly interested in showing data at large angles, so we didn't apply the 
necessary care to reduce the noise of the data at low angle (avoid convection, 
take larger statistical samples, extend the measurement over longer time spans,
and other technicalities). Furthermode, systematic errors at low angles can 
be also introduced by the fact that low angle data are obtained by collecting 
light nearly along the impinging beam direction, corresponding to 
light entering at large angles into the objective, where aberrations are often present.

\begin{figure}[h]
\begin{tabular}{cc}
\begin{picture}(120,120)(0,0)
\put(0,0){\includegraphics[scale=0.6]{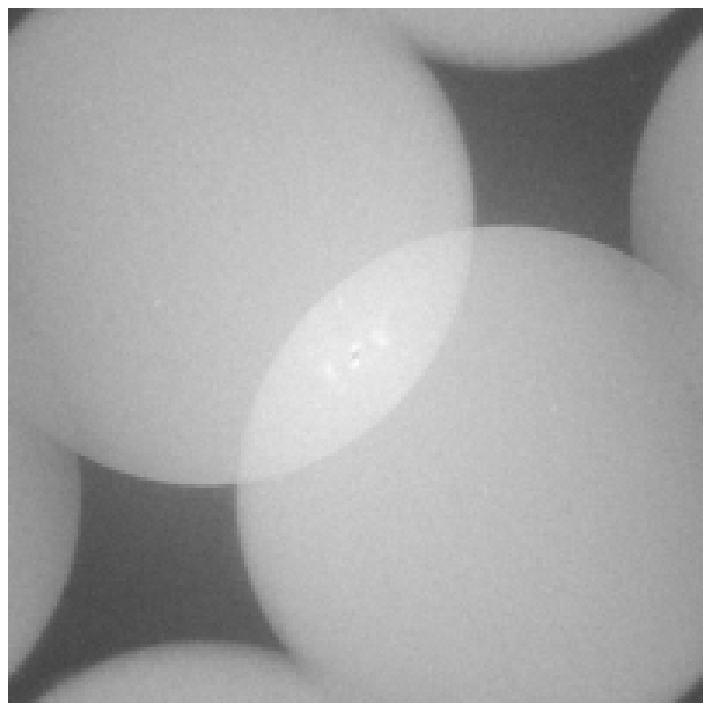}}
{\color{red} \path(60,60)(120,0)}
\end{picture}
&
\begin{picture}(120,120)(0,0)
\put(0,0){\includegraphics[scale=0.6]{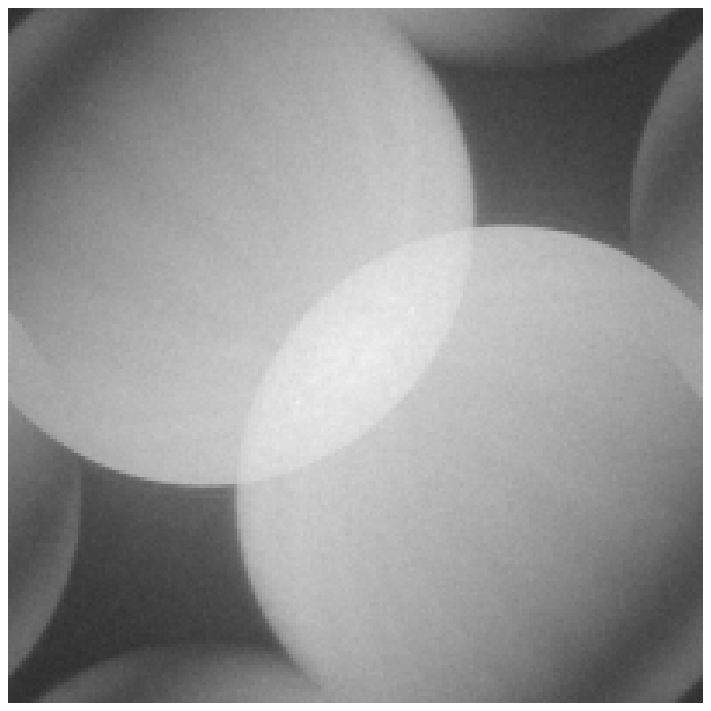}}
{\color{red} \path(60,60)(120,0)}
\end{picture}
\\
\includegraphics{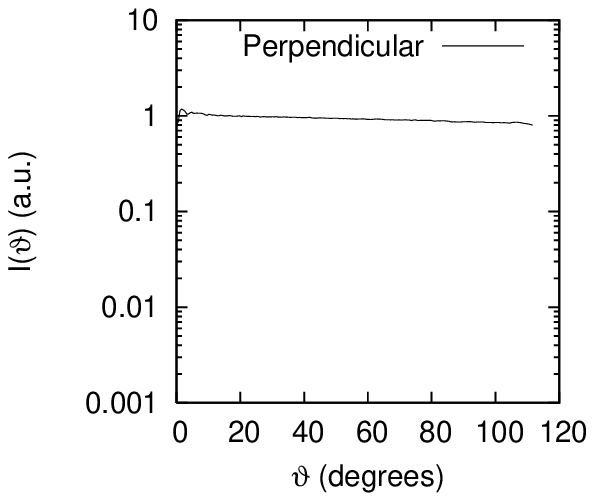}
&
\includegraphics{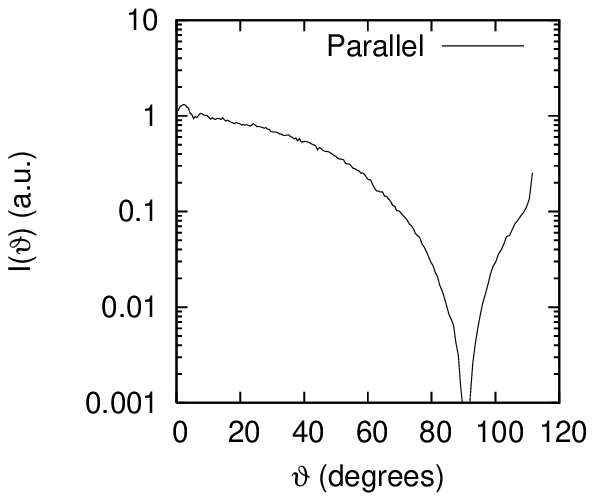}
\end{tabular}
\caption{Upper panels: Intensity Power Spectra IPS measured in two different polarization conditions. 
Lower panels: log-log polar plot of $I\left(\vartheta\right)$,
along the diagonal ($\varphi=0$, red line). 
Left column: perpendicular component of scattering.
Right column: parallel component of scattering.
The spectra, for the parallel case, 
shows dark bands representing the minimum of scattering
at $90^{\circ}$ along the polarization direction, 
due the dipolar scattering. Data obtained with sample B.
}
\label{fig_polarizzato}
\end{figure}

A further check of the capabilities of the technique has been
obtained by measuring the $S\left(\vartheta\right)$ with two
different polarizations of the impinging laser beam, i.e. parallel
and perpendicular to the azimuthal angle $\varphi=0^{\circ}$.
Fig.~\ref{fig_polarizzato} shows the results obtained for the 
colloidal sample B, for both parallel (right panels) and perpendicular
(left panels) components of the scattered light. 
The IPS are shown in the upper panels,
while the corresponding $I\left(\vartheta\right)$ are shown in the lower panels.
For the parallel case shown in the upper right panel,
the dark bands in the corners 
represent the minimum of scattering intensity
at $90^{\circ}$ along the polarization direction, a consequence of the
dipolar radiation of the scatterers, showing the usual factor $\cos^2(\vartheta)$. 
The same feature can be observed
in the corresponding $I\left(\vartheta\right)$ spectrum (lower right panel).

\begin{figure}[h]
\begin{tabular}{cc}
\includegraphics{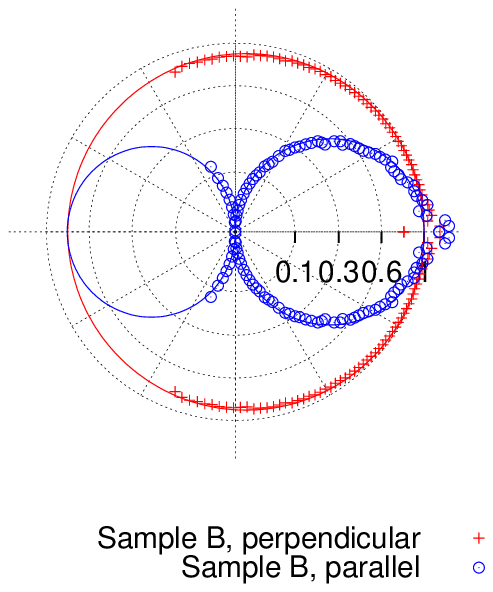}
&
\includegraphics{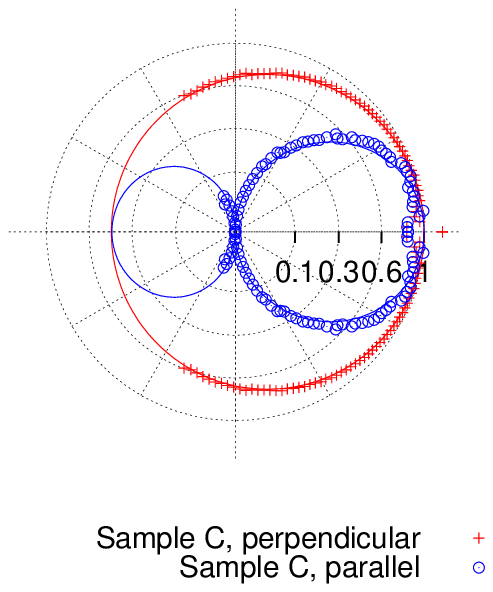}
\end{tabular}
\caption{Polar plots of the perpendicular and parallel components of the scattering intensity 
$I\left(\theta\right)$ measured as a function of the scattering angle $\vartheta$.
Data obtained with colloidal samples B (left panel) and C (right panel). 
Intensity has been mapped as a square root, in order to emphasize
the minima around $90^{\circ}$ of the parallel component. 
}
\label{fig_polare}
\end{figure}

Fig.~\ref{fig_polare} shows the measured 
parallel and perpendicular scattering components
$I\left(\vartheta\right)$, for two colloids B and C. Also in this case
the data are compatible with the Mie theory, and one can easily
detect the $90^{\circ}$ drop in the scattered intensity of the
parallel component (blue dots in Fig.~\ref{fig_polare}).
\begin{figure}[h]
\includegraphics{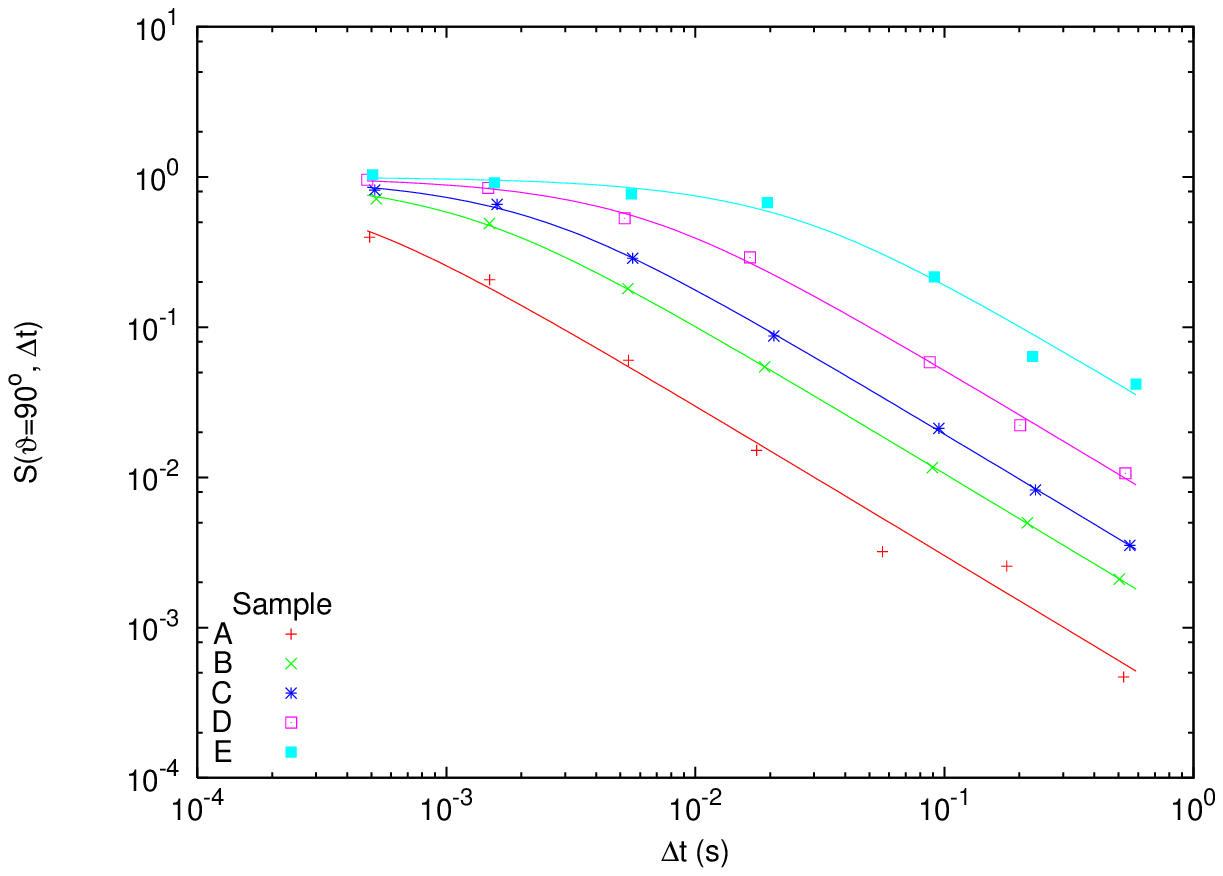}
\caption{Dynamic SINF measurements. 
Exposure-Time Dependent Spectra (ETDS), measured as a function of exposure time $\Delta t$,
for $\vartheta=90^{\circ}$. Data obtained with samples A, B, C, D, E.
Fitting lines are theoretical ETDS functions (see Eq.~\ref{eq_ETDf}), with decay time $\tau=1/DQ^2$, and
$D$ calculated with Stokes Einstein formula, using the values of nanoparticle 
diameters measured with dynamic SIFF and reported in Tab.~\ref{tab_samples}.
}
\label{fig_dinamico}
\end{figure}
Fig.~\ref{fig_dinamico} shows a log-log plot of the ETDS
measured as a function of exposure time $\Delta t$, for $\vartheta=90^{\circ}$,
for the different samples A, B, C, D, E.
Fitting lines are theoretical ETDS functions (see Eq.~\ref{eq_ETDf}), with decay time $\tau=1/DQ^2$, and
$D$ calculated with Stokes Einstein formula.
The data have been normalized so that the asymptotic value at short time is equal to 1.
We have used in the ETDS function a value of 
the nanoparticle diameter as measured with dynamic SIFF and reported in Tab.~\ref{tab_samples}.
The expected dependence $f\left(\Delta t/\tau\right)$ is recognized.
Also this experiment confirms that TLM allows detection of the scattered light at $90^{\circ}$.

\section{Conclusions.}

In the present paper, we have applied a new illumination geometry 
to a Scattering In the Near Field set up in order 
to increase the range of available scattering angles of the technique. 
The apparatus is based on a microscope equipped with a coherent source of
light not collinear with the objective optical axis (Tilted Laser Microscopy). 
This enables measuring both the static and dynamic spectra of
the sample up to backscattering angles
($110^{\circ}$), which was never achieved before with a SINF set up.
The improvement of the wave vector range over the
state-of-the-art \cite{ferri2004} is of a factor 7, and 
3.5 with respect to our experiments \cite{brogioli2009}.
We introduced for the first time a non-trivial analysis
algorithm necessary for extracting useful data.
The data processing takes advantages of the ETDS approach to get access to time
scales of the order of the camera minimum exposure time. The static
and dynamic data have been compared with calibrated
nanoparticles of various sizes in different polarization schemes.

\section{Acknowledgements.}

The authors are pleased to acknowledge fruitful discussion with F.
Croccolo, C. Lancellotti and financial funding from EU (projects Bonsai
LSHB--CT--2006--037639 and NAD CP--IP--212043--2).

\end{document}